# Thermal conductivity of MoS$_2$ polycrystalline nanosheets

M. Sledzinska[1,*], B. Graczykowski[1], M. Placidi[2], D. Saleta Reig[1,6], A. El Sachat[1], J.S. Reparaz[1], F. Alzina[1], S. Roche[1,5], B. Mortazavi[3], R. Quey[4], and C. M. Sotomayor Torres[1,5]

[1] Catalan Institute of Nanoscience and Nanotechnology (ICN2), CSIC and The Barcelona Institute of Science and Technology, Campus UAB, Bellaterra, 08193 Barcelona, Spain

[2] Catalonia Institute for Energy Research (IREC) Jardín de les Dones de Negre 1, 08930, Sant Adrià del Besòs, Spain

[3] Advanced Materials Multiscale Modeling, Institute of Structural Mechanics, Bauhaus-Universität Weimar, Marienstr. 15, D-99423 Weimar, Germany

[4] École des Mines de Saint-Étienne, CNRS UMR 5307, 158 cours Fauriel, 42023 Saint-Étienne, Cedex 2, France

[5] Institució Catalana de Recerca i Estudis Avançats (ICREA), 08010 Barcelona, Spain

[6] Departamento de Física, Universidad Autónoma de Barcelona, 08193 Bellaterra, Spain

**Abstract**

We report a technique for transferring large areas of the CVD-grown, few-layer MoS$_2$ from the original substrate to another arbitrary substrate and onto holey substrates, in order to obtain free-standing structures. The method consists of a polymer- and residue-free, surface-tension-assisted wet transfer, in which we take advantage of the hydrophobic properties of the MoS$_2$. The methods yields better quality transferred layers, with fewer cracks and defects, and less contamination than the widely used PMMA-mediated transfer and allows fabrication of few-layer, fee-standing structures with diameters up to 100 µm. We report thermal conductivity measurements by means of contactless Raman thermometry on the so-fabricated samples. The measurements revealed a strong reduction in the in-plane thermal conductivity down to 0.5 Wm$^{-1}$K$^{-1}$. The results are explained using finite elements method simulations for a polycrystalline film.

**Introduction**

Transition metal dichalcogenides, such as MoS$_2$, are two-dimensional materials which have recently attracted a lot of attention due to their optical properties, such as the thickness-dependent electronic band-gap transition and large optical absorption, which covers almost the whole visible spectrum[1,2]. At the same time they show excellent room-temperature carrier mobility with a high on-off ratio making them perfect candidates for nano-electronic[3-5] and -optoelectronic[6-8] applications.

---

[*] Corresponding author.
E-mail address: *marianna.sledzinska@icn.cat*

Thermal properties of $MoS_2$ have not been studied as intensively as optical and electronic ones, even though these properties are crucial for thermal management of any future 2D device. The thermal conductivity of the MoS2 crystals show strong anisotropy between the in-plane and cross-plane directions, reported to be 85-100 W/mK and 2 W/mK, respectively for the bulk (001) monocrystal[9].

Measurements of the in-plane thermal conductivity of a single layer, exfoliated single-crystalline $MoS_2$ using Raman thermometry by Yan et al. gave a value of 34.5 W/mK [10], while the thermal conductivity of a few-layer film prepared using a high-temperature vapour-phase method was of approximately 52 W/mK [11]. Similarly, Jo et al. found values between 44-55 W/mK for a few-layer $MoS_2$ using a suspended heater-thermometer structure [12].

In the case of poly-crystalline $MoS_2$, $WS_2$ and $WSe_2$ thin films, typically 50–150 nm thick, deposited by sputtering, the cross-plane thermal conductivity, measured by time domain thermoreflectance, was reported to be 10 times lower than for a bulk sample [13]. For thin $MoS_2$ films in particular, the thermal conductivity was measured for both orientations and found to be approximately 1.5 W/mK along basal planes and 0.25 W/mK across the basal planes [14]. These measurements demonstrate the importance of thermal boundary scattering as the limiting factor for thermal conductivity in nano-crystalline $MoS_2$ thin films.

Here we report a technique for transferring large areas of $MoS_2$ nanosheets from the original substrate to another arbitrary substrate and onto holey substrates to obtain free-standing structures. The method consists of a polymer- and residue-free, surface-tension-assisted wet transfer and takes advantage of the hydrophobic properties of the $MoS_2$. The method yields better quality transferred layers, with fewer cracks and defects, and less contamination than the widely used PMMA-mediated transfer and allows fabrication of few-layer, free-standing structures with diameters up to 100 μm.

Thermal conductivity measurements were performed using non-contact, 2-laser Raman thermometry, which has proven to be a useful method for measuring thermal conductivity of free-standing nano-membranes [15]. Compared to other studies, we have improved the measurement accuracy by using large area membranes (100 μm compared to 1.2 μm in ref. [10]) which reduces the influence of heat flow to the substrate. Moreover, the circular shape assures uniform heat flow over the membrane. We measured directly the laser power absorbed by the suspended samples under study. Also all the measurements are performed in vacuum to reduce convective cooling. Understanding the thermal properties of $MoS_2$ can give an insight on the thermal transport in ultra-thin semiconducting films, especially taking into account the grainy structure of polycrystalline materials.

**Fabrication and Structural Characterization**

In order to obtain free-standing films we have transferred 5 nm $MoS_2$ layers, fabricated via thin-film conversion technique, from the original Si/ $SiO_2$ substrate onto pre-patterned 200 nm thick $Si_3N_4$ membranes (Norcada Inc.). Various approaches to transferring thin films from one substrate to another have been reported in the literature, typically using a carrier polymer, such as polystyrene or PMMA [16, 17]. However, the carrier polymer can leave residues on the thin film, which could influence the thermal conductivity measurements. Therefore in this work we report an etching- and polymer-free, surface tension-assisted wet transfer method, enabling the fabrication of freely suspended, drum-like structures with diameters up to 100 μm.

The scheme of the method is shown in Fig. 1a). In this simple, three-step process we take advantage of the MoS$_2$ hydrophobic properties, which are in contrast to the hydrophilic properties of the growth substrate (SiO$_2$). In the first step, the CVD-grown material is directly submerged in DI water in order detach the thin film material from the substrate. The different surface energies drive water molecules to penetrate underneath the film and the MoS$_2$ film, which is left floating on the water surface. Holey Si$_3$N$_4$ membranes are used to scoop the floating thin film from the water surface and the samples are dried on a hotplate at 100 C. With this technique successful transfers were performed from SiO$_2$ to holey membranes with high free-standing rates: almost 90% for 10 μm diameter holes and approximately 40% for 60, 80 and 100 μm diameter holes.

Atomic force microscopy images showed a uniform MoS$_2$ film with thickness of 5 nm (Fig 1c)) and the average roughness of approx. 1 nm, which was expected, given the polycrystalline nature of the film.

High resolution TEM images confirmed the polycrystalline nature of the sample and the grain-size distribution statistics obtained, from which the average grain size was found to be between 4 and 6 nm, as shown in Fig 1 d) and e). The TEM diffraction pattern was also typical for powders and poly-crystals and reveals the presence of various crystallographic orientations in the film.

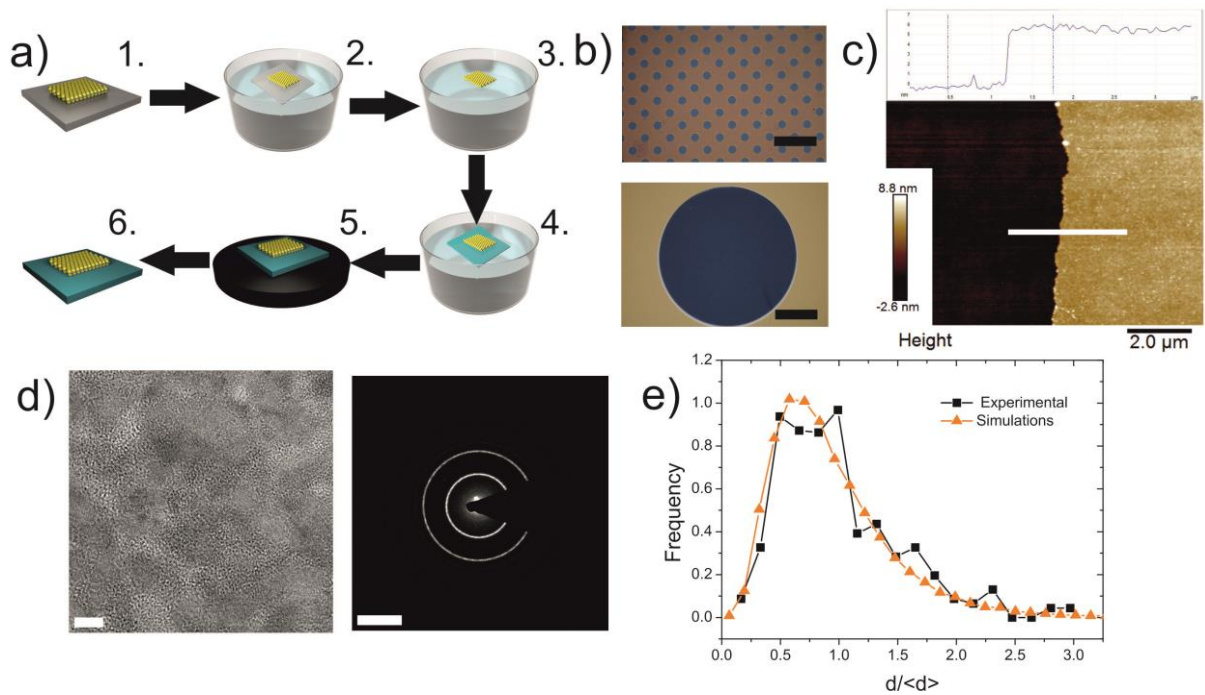

Fig.1 a) Schematic of the polymer-free, surface-energy-assisted transfer process. b) Optical micrographs of the free-standing MoS$_2$ membranes fabricated via polymer-free transfer. Scale bars correspond to 50 μm and 25 μm, top and bottom images, respectively. c) AFM image of the edge of the MoS$_2$ thin film after transfer to a Si$_3$N$_4$ substrate. Inset: Section of the marked area. d) TEM image of the MoS$_2$ sample and the diffraction pattern, showing clear polycrystalline structure. Scale bars are 5 nm and 5 nm$^{-1}$, respectively. e) Histogram of the grain sizes present in the sample (red). The majority of the grains have the size of 3-6 nm. The blue line represents the fitting of the histogram used in FEM thermal conductivity simulations.

**Results and discussion**

Raman scattering of bulk transition metal dichalcogenides has been studied since the 1970s, resulting in the identification of two principal first-order modes: $A_{1g}$ (410 cm$^{-1}$) and $E_{2g}$ (382 cm$^{-1}$) [18]. The Raman peak position of these two modes is temperature dependent presenting thermal softening, i.e., with increasing temperature we observe a red-shift in both $A_{1g}$ and $E_{2g}$ modes. The variation of Raman scattering spectrum with temperature is shown in Fig. 2a).

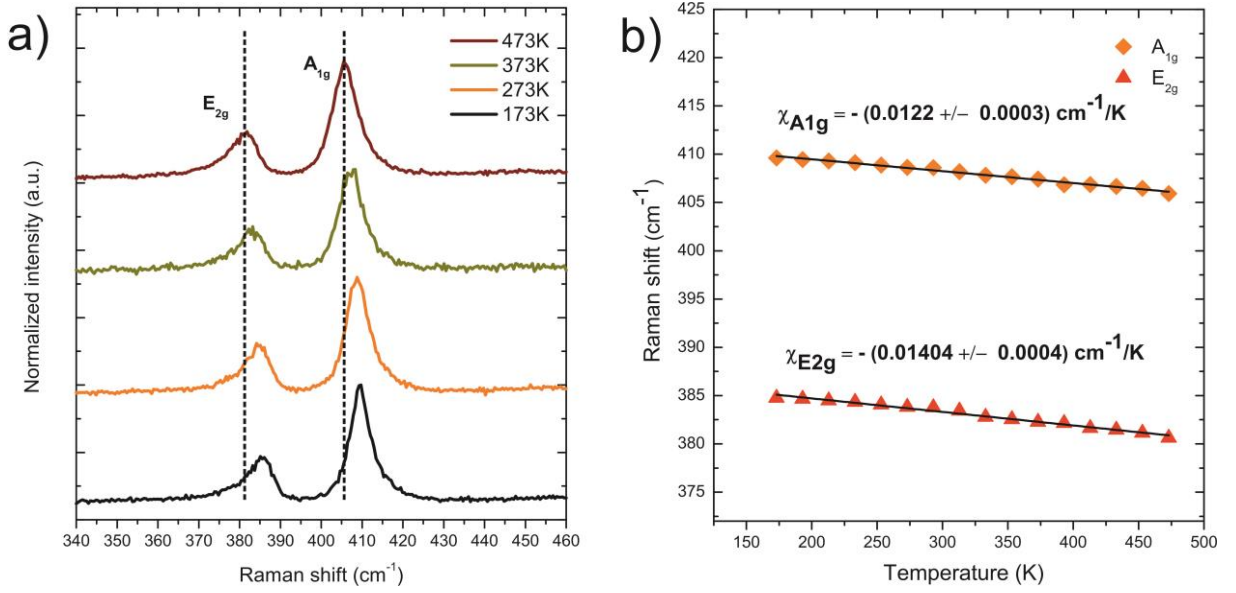

Fig. 2 Temperature-dependent Raman spectra of MoS$_2$ a) Spectra recorded at different temperatures. b) Temperature dependence of the frequencies of Raman active $A_{1g}$ and $E_{2g}$ modes.

The frequencies of the active $A_{1g}$ and $E_{2g}$ modes as a function of temperature are plotted in Fig. 2b). The data was fitted to the linear equation:

$$\omega(T) = C + \chi T \qquad (1)$$

The values obtained were $\chi_{E2g}$ = -1.4x10$^{-2}$ cm$^{-1}$/K for $E_{2g}$ mode and $\chi_{A1g}$ = -1.22x10$^{-2}$ cm$^{-1}$/K for $A_{1g}$ mode, which are comparable to the bulk [19] and nanosheets values reported in the literature [11].

To measure the thermal conductivity of the nanosheets we used the 2-laser Raman thermometry, shown schematically in Fig. 3a). The basic idea of this method relies on the use of two lasers: one in the centre of the membrane which serves as a steady-state heater and another, low power, probing laser, which measures the Raman shift along a line on the membrane. The details of the technique are explained in the ref [15]. In the case of a 2-dimensional membrane the Fourier heat equation has the following solution:

$$\frac{dT}{d\ln(r)} = -\frac{P_{abs}}{2\pi d\kappa} \qquad (2)$$

Where *r* is the relative position on the membrane with respect to the position of the heating laser, $P_{abs}$ is the absorbed power and *d* is the membrane thickness. Using the correct value of the absorbed power is crucial as it directly influences the thermal conductivity calculated from Eq. (2). We have obtained the value of $P_{abs}$ by subtracting from the incident laser power the power transmitted and reflected by the $MoS_2$ nanosheets.

The experimental data of the relative Raman shift of the $A_{1g}$ mode as a function of position *r* is shown in Fig 3c). The maximum Raman shift if observed at the centre of the membrane, where the heating laser is placed. Using the $\chi_{A1g}$ coefficient the Raman shift is converted into a thermal field, to give the temperature decay over the membrane diameter. Figure 3c) displays the temperature decay in logarithmic scale presenting a linear dependence. Therefore, the thermal conductivity can be obtained according to eq. (2) from the inverse of the slope of the linear adjustment to the experimental data. The thermal conductivity obtained with this procedure was found to be 0.5 ±0.002 W/mK.

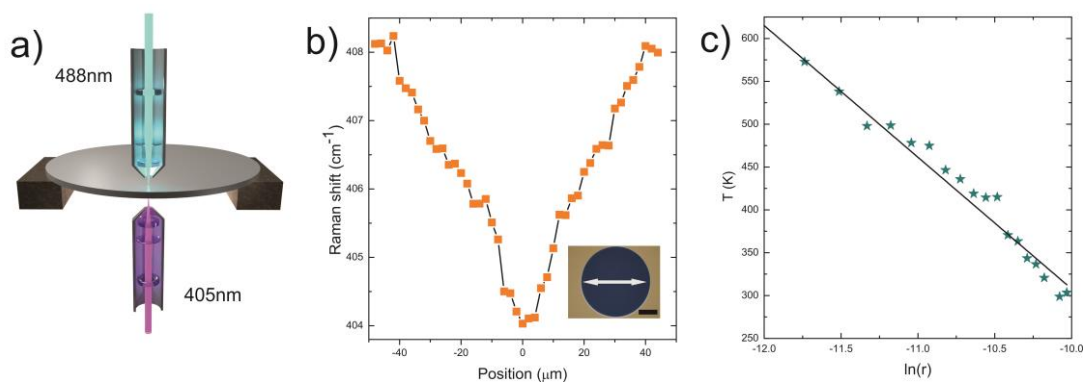

Fig. 3 a) Measurement scheme of the two-laser Raman thermometry set-up. b) Raman shift of the $A_{1g}$ mode as a function of position. Inset: Optical microscope image of the measured sample. Arrow indicates the scan direction. The scale bar is 25 µm. c) Temperature profiles as a function of the ln(r), where r is the distance from the center to the edge of the membrane in meters. Solid line represents linear fitting of the experimental points.

We simulated the reduction of the thermal conductivity using finite element modelling (FEM). A recent study showed the validity of the continuum model to describe the thermal conductivity of nanocrystalline 2D films[20]. In the continuum modelling developed here, we reconstruct a polycrystalline sample consisting of 3000 grains. The grain size distributions for experimental and reconstructed samples are plotted on Fig. 1e), which clearly confirms that the reconstructed model with a lognormal distribution is an accurate representative of the experimental sample.

In our continuum modelling, we assumed that all grain boundaries present an effective thermal contact conductance. This effective conductance was estimated by fitting the finite element result to the experimental measurement for thermal conductivity of a sample with a grain size of 5 nm. In our modelling, we included two highly conductive strips at the two ends of polycrystalline film [21]. We then applied a constant inward and outward heat fluxes (q) on the outer surfaces of these strips. The steady-state temperature difference along the structure, ΔT, was then computed, which correlates to the sample effective thermal conductivity, $\kappa_{eff}$, by $\kappa_{eff} = \frac{q}{\Delta T/L}$, where *L* is the sample length.

In Fig. 4, the calculated steady-state temperature and heat-flux profiles of a sample with an average grain size of 5 nm are illustrated. In this calculation, we assume a thermal conductivity of 100 W/mK for $MoS_2$ films [22]. Our continuum simulations reveal that for a sample with equivalent grain size of 5 nm, the temperatures inside individual grains are visibly constant (Fig. 4a) and the temperature changes occur across the grain boundaries. In this case, the grains are not equally involved in the heat flux transfer and the majority of heat flux is transferred by the large grains that are percolating together (Fig. 4b). This finding implies that for a sample with a grin size of 5 nm the grain boundary conductance is the factor that dominates the heat transfer. Consequently, the heat-flux is mainly transferred along percolations paths in which there exist minimum grain boundaries. Our continuum calculation predicts an effective grain boundary conductance of 88.2 MW/m$^2$K for $MoS_2$ grain boundaries. We also performed our calculations by assuming a thermal conductivity of 34.5 W/mK for single-crystalline $MoS_2$ sheets [10]. For this case, we found a grain boundary conductance of 86.4 MW/m$^2$K, which is remarkably close to our earlier prediction. Thus, our investigation suggests an effective grain boundary conductance of 87.5±1.5 MW/m$^2$K for $MoS_2$ films. Worthy to note that the grain boundary conductance of graphene was found experimentally to be in a range between 260-2250 MW/m$^2$K [23].

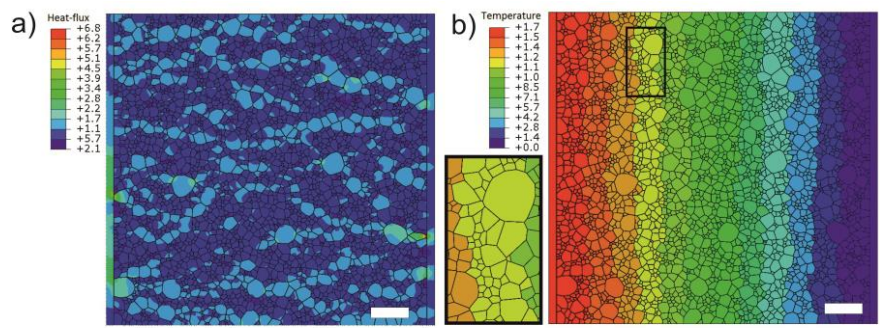

Fig 4. a) Simulated heat flux in the sample. Scale bar corresponds to 30 nm. b) Temperature distribution in the sample. Scale bar 30 nm. Inset: Detail of the marked region.

**Conclusions**

We have investigated the thermal properties of $MoS_2$ polycrystalline nanosheets using non-contact 2-laser Raman thermometry. We have observed a strong reduction in the thermal conductivity of the nanosheets, due to the scattering on grain boundaries. Both experimental and theoretical results included in this work give an insight on the phonons in polycrystalline nanomaterials. The possibility of tailoring thermal conductivity by controlling the grainsizes in the polycrystalline materials offers multiple opportunities for future devices. Especially the strong reduction of thermal conductivity reported in this work for thin nanosheets can find use in 2D energy devices [24, 25]. However, the detailed study to optimize grainsize, thickness and doping level remains to be performed.

**Experimental section**

*Material growth and characterization*

The synthesis of the MoS$_2$ nanosheets was achieved via thin-film conversion technique. A thin (2 nm) Mo layer was deposited by DC-magnetron sputtering on 100 mm Si/SiO$_2$ substrates and reactively annealed in a graphite box under sulfur containing atmosphere at 600ºC during 30 minutes. The main advantage of this method is the easy control of the thickness, down to a few nanometers and the possibility to scale it up to 4 inch wafers. However, the resulting material is polycrystalline and the sub-nanometer thickness is difficult to achieve, even though some recent works indicate that this can be overcome in order to achieve high quality films [11, 26].

Atomic force microscopy topographies were done using Asylum MFP-3D.

High-resolution imaging of structure and morphology of the samples was performed using FEI Tecnai F20 in TEM and STEM modes. Grain distribution analysis was performed using Digital Micrograph and ImageJ software.

*Raman thermometry*

The temperature dependent Raman scattering measurements were performed in *Lincam* cryostat under vacuum, varying the temperature between 100 and 475 K, and using 488 nm laser line (Spectra Physics, model 2018) to obtain the spectra.

The two laser Raman thermometry measurements were performed using a 488 nm probing laser (Spectra Physics, model 2018) and 405 nm heating laser (Cobolt). The power of the probing laser was kept to the minimum (2 µW) in order to avoid extra heating. The heating laser power was set to 36 µW, which resulted in 0.026 µW of P$_{abs}$. All the measurements were performed at room temperature in *Lincam* cryostat at 10$^{-3}$ mBar pressure on MoS$_2$ membranes with diameter of 80 and 100 µm.

*Finite element modelling*

The polycrystalline sheet was modelled as a Laguerre tessellation, using a new algorithm developed in the Neper software package [27]. A Laguerre tessellation of a domain of space, D, is constructed from a set of seeds, S$_i$, of positions, **x**$_i$, and weights, w$_i$ (i = 1, …, N). Each seed is assigned a Laguerre cell, C$_i$, defined by

$$C_i = \{P(x) \in D, d(P,S_i)^2 - w_i^2 < d(P,S_j)^2 - w_j^2 \forall i \neq j\} \quad (4)$$

where 'd' is the Euclidean distance. The set of cells form a partition of D representing the polycrystal. Various methodologies have been proposed to determine seed attributes to match specific polycrystal properties, the most popular of which is dense spherical packing. Here, we rather adopted an optimization approach for which the seeds attributes are set by minimization of an objective function, A,

$$A = \int_{-\infty}^{+\infty} [F^*(x) - F(x)]^2 \, dx \quad (5)$$

where F* and F stand for the numerical and experimental cumulative density functions of the grain size distributions, respectively. Function A reaches the minimal value of 0 when F* reaches F. (In this work, F was taken as a lognormal distribution with a standard deviation / average ratio of 0.52.)

The non-linear optimization problem was solved using NLOpt's Subplex algorithm. The microstructure was then meshed into triangular elements using standard algorithms and interface nodes were duplicated using an in-house Python script. Finally, the thermal conductivity problem was solved using the finite element code Abaqus.

**Acknowledgments**

The authors acknowledge the financial support from the FP7 FET Energy Project MERGING (Grant no. 309150); NMP QUANTIHEAT (Grant no. 604668), the Spanish MICINN projects nanoTHERM (Grant no. CSD2010-0044); TAPHOR (MAT2012-31392) and the Severo Ochoa program (Grant SEV-2013-0295).